\documentclass[english,pre,a4paper,aps,reprint]{revtex4-2}
\usepackage[T1]{fontenc}
\usepackage[latin9]{inputenc}
\usepackage{amsmath}
\usepackage{amssymb}
\usepackage{graphicx}
\usepackage{babel}
\begin{document}
\title{Three components of stochastic entropy production associated with
the quantum Zeno and anti-Zeno effects }
\author{Ashutosh Kinikar and Ian J. Ford }
\affiliation{Department of Physics and Astronomy, University College London, Gower
Street, London WC1E 6BT, U.K.}
\begin{abstract}
We investigate stochastic entropy production in a two-level quantum
system that performs Rabi oscillations while undergoing quantum measurement
brought about by continuous random disturbance by an external measuring
device or environment. The dynamics produce quantum Zeno and anti-Zeno
effects for certain measurement regimes, and the stochastic entropy
production is a measure of the irreversibility of the behaviour. When
the strength of the measurement disturbance is time-dependent, the
stochastic entropy production separates into three components. Two
represent relaxational behaviour, one being specific to systems represented
by coordinates that are odd under time reversal symmetry, and a third
characterises the nonequilibrium stationary state arising from breakage
of detailed balance in the dynamics. The study illustrates how the
ideas of stochastic thermodynamics may be applied in similar ways
to both quantum and classical systems.
\end{abstract}
\maketitle

\section{Introduction}

Entropy quantifies subjective uncertainty in the configuration of
a system and it can be argued that similar applications of this concept
should apply in both classical and quantum mechanics, where configurations
are described by phase space coordinates and by elements of a density
matrix, respectively. The effective stochastic dynamics of such variables
brought about by coupling the system to a coarse grained environment
will increase the configurational uncertainty of the world (the system
together with its environment) as time passes. Such a loss of information
is often manifested in the dispersal of energy and matter or the loss
of correlations: consequences of the chaotic nature of the underlying
deterministic dynamics but nevertheless captured by stochastic modelling.
This is the content of the second law of thermodynamics \citep{fordbook}.

The aim of this paper is to compute the stochastic entropy production
and hence loss of information when a simple quantum system undergoing
Rabi oscillations is subjected to continuous measurement of two non-commuting
observables \citep{jacobs2014quantum}. The system exhibits Zeno and
anti-Zeno effects \citep{misra1977a} depending on the relative strengths
of the two measurement processes. It is of particular interest to
consider the division of the stochastic entropy production into three
components when the strength of measurement is time-dependent \citep{SpinneyFord12a,SpinneyFord12b}.
Each component describes an aspect of the nonequilibrium, irreversible
behaviour of the system.

In Section \ref{sec:Stochastic-dynamics} we derive Markovian stochastic
differential equations, or Itô processes, that describe the evolution
of the system. We examine Zeno and anti-Zeno effects where the mean
rate of change of a system coordinate is reduced or increased, respectively,
when measurement is made more intense. The nature of the three components
of stochastic entropy production for time-dependent measurement of
one of the observables is discussed in Section \ref{sec:Stochastic-entropy-production}.
We give our conclusions in Section \ref{sec:Conclusions}.

\section{Stochastic dynamics\label{sec:Stochastic-dynamics}}

The reduced density matrix $\rho$ is a specification of the state
of an open quantum system and under certain conditions of coupling
to the environment its evolution can be modelled using a stochastic
Lindblad equation:
\begin{align}
d\rho & =-i[H,\rho]dt+\sum_{i}\left(c_{i}\rho c_{i}^{\dagger}-\frac{1}{2}\rho c_{i}^{\dagger}c_{i}-\frac{1}{2}c_{i}^{\dagger}c_{i}\rho\right)dt\nonumber \\
 & \qquad+\left(\rho c_{i}^{\dagger}+c_{i}\rho-C_{i}\rho\right)dW_{i},\label{two pairs stochastic lindblad}
\end{align}
with $C_{i}={\rm Tr}\left((c_{i}+c_{i}^{\dagger})\rho\right)$, where
$H$ is the system Hamiltonian. The Lindblad operators $c_{i}$ represent
the modes of interaction between the system and the environment, and
the $dW_{i}$ are a set of independent Wiener increments \citep{matos2022,Clarke23,Dexter23}.
This framework is a form of quantum state diffusion, where evolution
of $\rho$ is continuous, without jumps \citep{percival1998}.

We consider a two-level bosonic system represented by $\rho=\frac{1}{2}\left(\mathbb{I}+\boldsymbol{r}\cdot\boldsymbol{\sigma}\right)$,
where $\boldsymbol{r}$ is the coherence or Bloch vector and $\sigma_{k}$
are the Pauli matrices, with $H=\epsilon\sigma_{z}$, $c_{1}=\alpha_{x}\sigma_{x}$
and $c_{2}=\alpha_{y}\sigma_{y}$. The dynamics describe a system
that performs Rabi oscillations in the expectation values $r_{x}={\rm Tr}(\sigma_{x}\rho)$
and $r_{y}={\rm Tr}(\sigma_{y}\rho)$ when isolated, but which departs
stochastically from such regular behaviour when the coupling coefficients
$\alpha_{x}$ and $\alpha_{y}$ are non-zero. The situation is similar
to a two-level system undergoing the measurement of one observable,
studied previously \citep{Walls22}.

The Lindblads $c_{1}$ and $c_{2}$ tend to drive the system towards
eigenstates of $\sigma_{x}$ and $\sigma_{y}$, respectively, and
their use in Eq. (\ref{two pairs stochastic lindblad}) can be regarded
as an implementation of the continuous, simultaneous measurement of
these two system observables \citep{Clarke23}. The coefficients $\alpha_{x}$
and $\alpha_{y}$ are \emph{measurement} strengths, since increasing
$\alpha_{x}$ while $\alpha_{y}$ is held constant brings about a
greater concentration of the pdf in the vicinity of the eigenstates
of $\sigma_{x}$, and vice versa.

The dynamics of the components of $\boldsymbol{r}$ can be expressed
as Itô processes:
\begin{align}
dr_{x} & \!=\!-2\left(\epsilon r_{y}+\alpha_{y}^{2}r_{x}\right)\!dt+2\alpha_{x}\left(1-r_{x}^{2}\right)\!dW_{x}\!-\!2\alpha_{y}r_{x}r_{y}dW_{y}\nonumber \\
dr_{y} & \!=\!2\left(\epsilon r_{x}-\alpha_{x}^{2}r_{y}\right)\!dt-2\alpha_{x}r_{x}r_{y}dW_{x}+2\alpha_{y}\left(1-r_{y}^{2}\right)\!dW_{y}\nonumber \\
dr_{z} & \!=\!-2r_{z}\left(\alpha_{x}^{2}+\alpha_{y}^{2}\right)\!dt-2r_{z}\left(\alpha_{x}r_{x}dW_{x}+\alpha_{y}r_{y}dW_{y}\right),\label{eq: 3sdes}
\end{align}
where $dW_{x}$ and $dW_{y}$ are Wiener increments.

We consider a (pure) state denoted by $r_{z}=0$, $r_{x}=\sin\phi$
and $r_{y}=\cos\phi$. The coherence vector lies in the equatorial
plane of the Bloch sphere and its rotation about the $r_{z}$ axis
is specified by an azimuthal angle $\phi=\tan^{-1}(r_{x}/r_{y})$.
The stochastic evolution of $\phi$ can be derived from Eq. (\ref{eq: 3sdes})
using Itô's lemma:
\begin{align}
d\phi & =\left(2\epsilon-\left(\alpha_{x}^{2}-\alpha_{y}^{2}\right)\sin2\phi\right)dt\nonumber \\
 & -2\alpha_{x}\sin\phi\,dW_{x}+2\alpha_{y}\cos\phi\,dW_{y}\nonumber \\
= & \left(2\epsilon-\left(\alpha_{x}^{2}-\alpha_{y}^{2}\right)\sin2\phi\right)dt\nonumber \\
 & -2\left(\alpha_{x}^{2}\sin^{2}\phi+\alpha_{y}^{2}\cos^{2}\phi\right)^{1/2}dW,\label{eq:dphi}
\end{align}
where $dW$ is also a Wiener increment. The dynamics produce a linear
increase in $\phi$ with time when the system is isolated ($\alpha_{x}=\alpha_{y}=0$).
This drift is distorted by random disturbances when the system is
coupled to the environment, here regarded as a measuring device. The
associated Fokker-Planck equation for the probability density function
(pdf) $p(\phi,t)$ is
\begin{align}
\frac{\partial p(\phi,t)}{\partial t} & =-\frac{\partial J}{\partial\phi},\label{eq:FPE}
\end{align}
where the probability current is
\begin{align}
J & =\left(2\epsilon-\left(\alpha_{x}^{2}-\alpha_{y}^{2}\right)\sin2\phi\right)p(\phi,t)\nonumber \\
 & -2\frac{\partial}{\partial\phi}\left(\alpha_{x}^{2}\sin^{2}\phi+\alpha_{y}^{2}\cos^{2}\phi\right)p(\phi,t).\label{eq:J}
\end{align}

The situation with equal non-zero measurement strengths $\alpha_{x}=\alpha_{y}=\alpha\ne0$
is described by $d\phi=2\epsilon dt-2\alpha dW$. The system then
evolves towards a stationary state with $p_{{\rm st}}(\phi)=(2\pi)^{-1}$
and $J=\epsilon/\pi$. When $\alpha_{x}\ne\alpha_{y}$, there is also
a stationary state with constant $J$ but characterised by a nonuniform
pdf. These are nonequilibrium situations with consequent stochastic
entropy production, which we investigate in the next section.

\begin{figure}
\begin{centering}
\includegraphics[width=1\columnwidth]{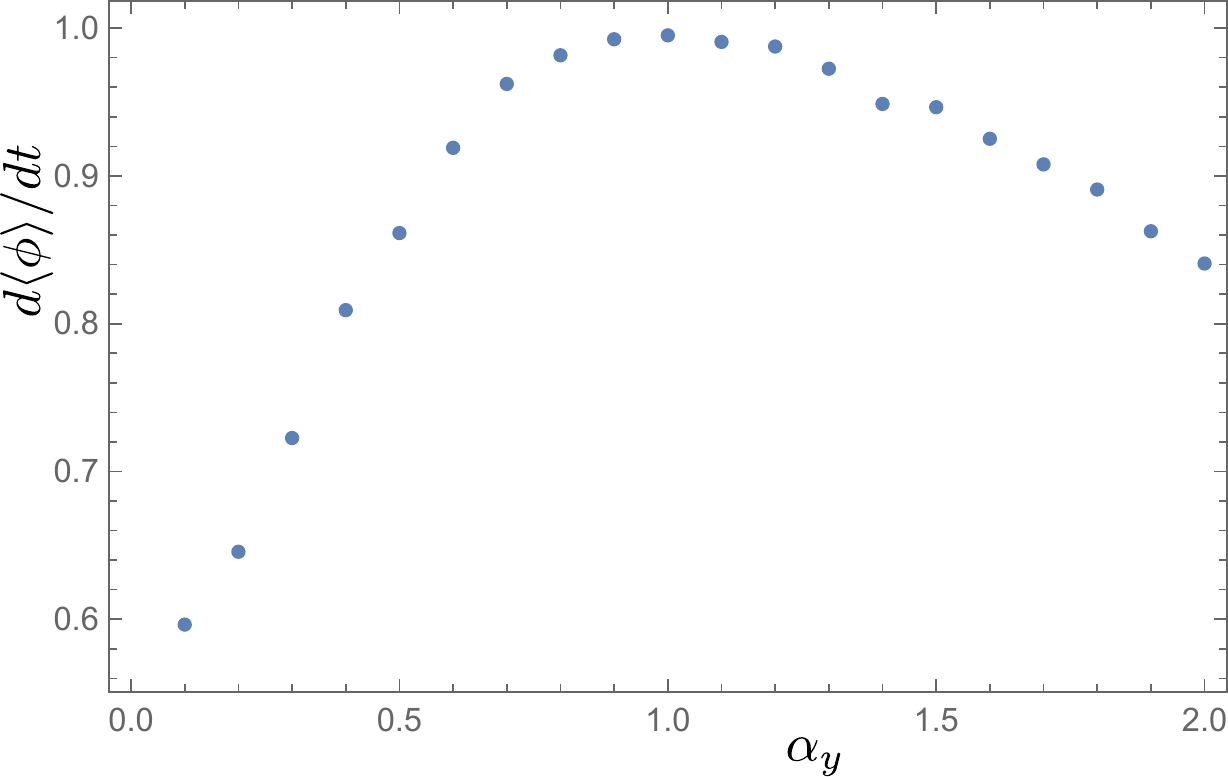}
\par\end{centering}
\caption{Zeno and anti-Zeno effects are apparent in the mean rate of Rabi oscillations,
arising from changing the measurement strength $\alpha_{y}$ for $\alpha_{x}=1$
and $\epsilon=1/2$. The average is obtained over a time interval
$\Delta t=10^{5}$ for each value of $\alpha_{y}$ and the error in
the mean is estimated to be less than 0.01. \label{fig:Zeno-and-anti-Zeno}}

\end{figure}

We consider the dependence of the mean rate of change of $\phi$ on
the measurement strengths $\alpha_{x}$ and $\alpha_{y}$. The mean
of $\phi$ evolves according to
\begin{equation}
\frac{d\langle\phi\rangle}{dt}=2\epsilon-\left(\alpha_{x}^{2}-\alpha_{y}^{2}\right)\langle\sin2\phi\rangle,\label{eq: dphi/dt}
\end{equation}
where the angled brackets represent an average over the stochasticity.
Results from numerical simulations of Eq. (\ref{eq:dphi}) are given
in Figure \ref{fig:Zeno-and-anti-Zeno} for $\epsilon=1/2$, $\alpha_{x}=1$
and a range of values of $\alpha_{y}$. The Zeno effect operates for
$\alpha_{y}>\alpha_{x}$; a slowing of the average evolution of the
system as the strength of measurement $\alpha_{y}$ is increased at
a constant $\alpha_{x}$ \citep{Walls22}. There is also an anti-Zeno
effect for $\alpha_{y}<\alpha_{x}$, where the mean evolution is speeded
up when $\alpha_{y}$ is increased. For $\alpha_{x}=\alpha_{y}$ and
$p_{{\rm st}}(\phi)=(2\pi)^{-1}$ the average of $\sin2\phi$ vanishes,
and $d\langle\phi\rangle/dt=2\epsilon$: the effects of the two measurement
processes on the mean rate of Rabi oscillation then cancel each other
out, somewhat counter-intuitively.

\section{Stochastic entropy production\label{sec:Stochastic-entropy-production}}

We now consider the stochastic thermodynamics associated with the
dynamics
\begin{equation}
d\phi=\left(A^{{\rm rev}}(\phi)+A^{{\rm irr}}(\phi)\right)dt+B(\phi,t)dW,\label{dynamics SDE}
\end{equation}
where the terms involving $A^{{\rm rev}}$ and $A^{{\rm irr}}$ represent
deterministic rates of change of $\phi$ that satisfy and violate
time reversal symmetry, respectively. The stochastic entropy production
is given by \citep{SpinneyFord12b}
\begin{eqnarray}
d\Delta s_{{\rm tot}} & = & -d\ln p(\phi,t)+\frac{A^{{\rm irr}}}{D}d\phi-\frac{A^{{\rm rev}}A^{{\rm irr}}}{D}dt+\frac{dA^{{\rm irr}}}{d\phi}dt\nonumber \\
 &  & -\frac{dA^{{\rm rev}}}{d\phi}dt-\frac{1}{D}\frac{\partial D}{\partial\phi}d\phi+\frac{(A^{{\rm rev}}-A^{{\rm irr}})}{D}\frac{\partial D}{\partial\phi}dt\nonumber \\
 &  & -\frac{\partial^{2}D}{\partial\phi^{2}}dt+\frac{1}{D}\left(\frac{\partial D}{\partial\phi}\right)^{2}dt,\label{dstot expression}
\end{eqnarray}
where $D(\phi,t)=\frac{1}{2}B(\phi,t)^{2}$. For dynamics that possess
an equilibrium state (a stationary state with vanishing probability
current $J$) characterised by a pdf $p_{{\rm st}}(\phi)$, Eq. (\ref{dstot expression})
reduces to the simpler expression $d\Delta s_{{\rm tot}}=-d\ln\left(p(\phi,t)/p_{{\rm st}}(\phi)\right)$,
showing explicitly how stochastic entropy production can arise from
a statistical deviation from equilibrium. The system under consideration
here, however, does not possess an equilibrium state in general, but
instead a nonequilibrium stationary state with non-zero $J$.

For bosonic systems, the time reversal operation corresponds to taking
a complex conjugate of the density matrix. Thus the components $r_{x}$
and $r_{z}$ of the coherence vector are even and the component $r_{y}$
is odd under time reversal symmetry. This means that $\phi$ is also
odd and we deduce that $A^{{\rm rev}}=2\epsilon$ and $A^{{\rm irr}}=-\left(\alpha_{x}^{2}-\alpha_{y}^{2}\right)\sin2\phi$.
The diffusion coefficient is $D=2\left(\alpha_{x}^{2}\sin^{2}\phi+\alpha_{y}^{2}\cos^{2}\phi\right)$.
We take the coefficients $\alpha_{x}$ and $\alpha_{y}$ to be time-independent
(for now) and write $dA^{{\rm irr}}/d\phi=-2\left(\alpha_{x}^{2}-\alpha_{y}^{2}\right)\cos2\phi$,
$dD/d\phi=2\left(\alpha_{x}^{2}-\alpha_{y}^{2}\right)\sin2\phi$,
$d^{2}D/d\phi^{2}=4\left(\alpha_{x}^{2}-\alpha_{y}^{2}\right)\cos2\phi$,
and obtain

\begin{align}
 & d\Delta s_{{\rm tot}}=-d\ln p+\frac{9\left(\alpha_{x}^{2}-\alpha_{y}^{2}\right)^{2}\sin^{2}2\phi}{2\left(\alpha_{x}^{2}\sin^{2}\phi+\alpha_{y}^{2}\cos^{2}\phi\right)}dt\nonumber \\
- & 6\left(\alpha_{x}^{2}-\alpha_{y}^{2}\right)\cos2\phi dt+\frac{3\left(\alpha_{x}^{2}-\alpha_{y}^{2}\right)\sin2\phi}{\left(\alpha_{x}^{2}\sin^{2}\phi+\alpha_{y}^{2}\cos^{2}\phi\right)^{1/2}}dW.\label{eq: dstot}
\end{align}
 For $\alpha_{y}=0$, and hence measurement of $\sigma_{x}$ alone,
this reduces to
\begin{equation}
d\Delta s_{{\rm tot}}=-d\ln p+6\alpha_{x}^{2}\left(1+\cos^{2}\phi\right)dt+6\alpha_{x}\cos\phi\,dW,\label{dstot 2}
\end{equation}
and we conclude that in a stationary state, for a given value of $\alpha_{x}$,
the stochastic entropy production increases on average at a constant
rate given by
\begin{equation}
\frac{d\langle\Delta s_{{\rm tot}}\rangle}{dt}=6\alpha_{x}^{2}\left(1+\langle\cos^{2}\phi\rangle\right),\label{eq: mean dstot}
\end{equation}
since $\langle-d\ln p\rangle=dS_{G}=0$ in these circumstances, where
$S_{G}=-\int p\ln p\,d\phi$ is the Gibbs entropy. For larger $\alpha_{x}^{2}$,
the pdf becomes more concentrated in the region of $\phi=0$ and $\pi$,
corresponding to the eigenstates of $\sigma_{x}$ \citep{Walls22},
such that $\langle\cos^{2}\phi\rangle$ increases with $\alpha_{x}^{2}$
towards an upper limit of unity. Thus a increase in measurement strength
brings about a higher mean rate of production of stochastic entropy,
which can be associated intuitively with the increased Zeno slowing
down, on average, of the Rabi oscillations.

We now consider a situation where the measurement strength $\alpha_{x}$
is time-dependent. In such circumstances the stochastic entropy production
separates into three identifiable components \citep{SpinneyFord12b},
written
\begin{equation}
d\Delta s_{{\rm tot}}=d\Delta s_{1}+d\Delta s_{2}+d\Delta s_{3}.\label{eq: 13}
\end{equation}
The rate of change of the mean value of the first component may be
written in the form
\begin{equation}
\frac{d\langle\Delta s_{1}\rangle}{dt}=-\int\frac{\partial p}{\partial t}\ln\frac{p(\phi,t)}{p_{{\rm st}}^{\alpha_{x}}(\phi)}d\phi.\label{eq: 14}
\end{equation}
Evidently, this is a relaxational entropy production that vanishes
when the system is in a stationary state characterised by the pdf
$p_{{\rm st}}^{\alpha_{x}}(\phi)$ associated with a specified value
of $\alpha_{x}$.  Esposito and Van den Broeck denoted this component
the nonadiabatic entropy production \citep{adiabaticnonadiabatic0}.
Its mean rate of change can never be negative.

\begin{figure}
\begin{centering}
\includegraphics[width=1\columnwidth]{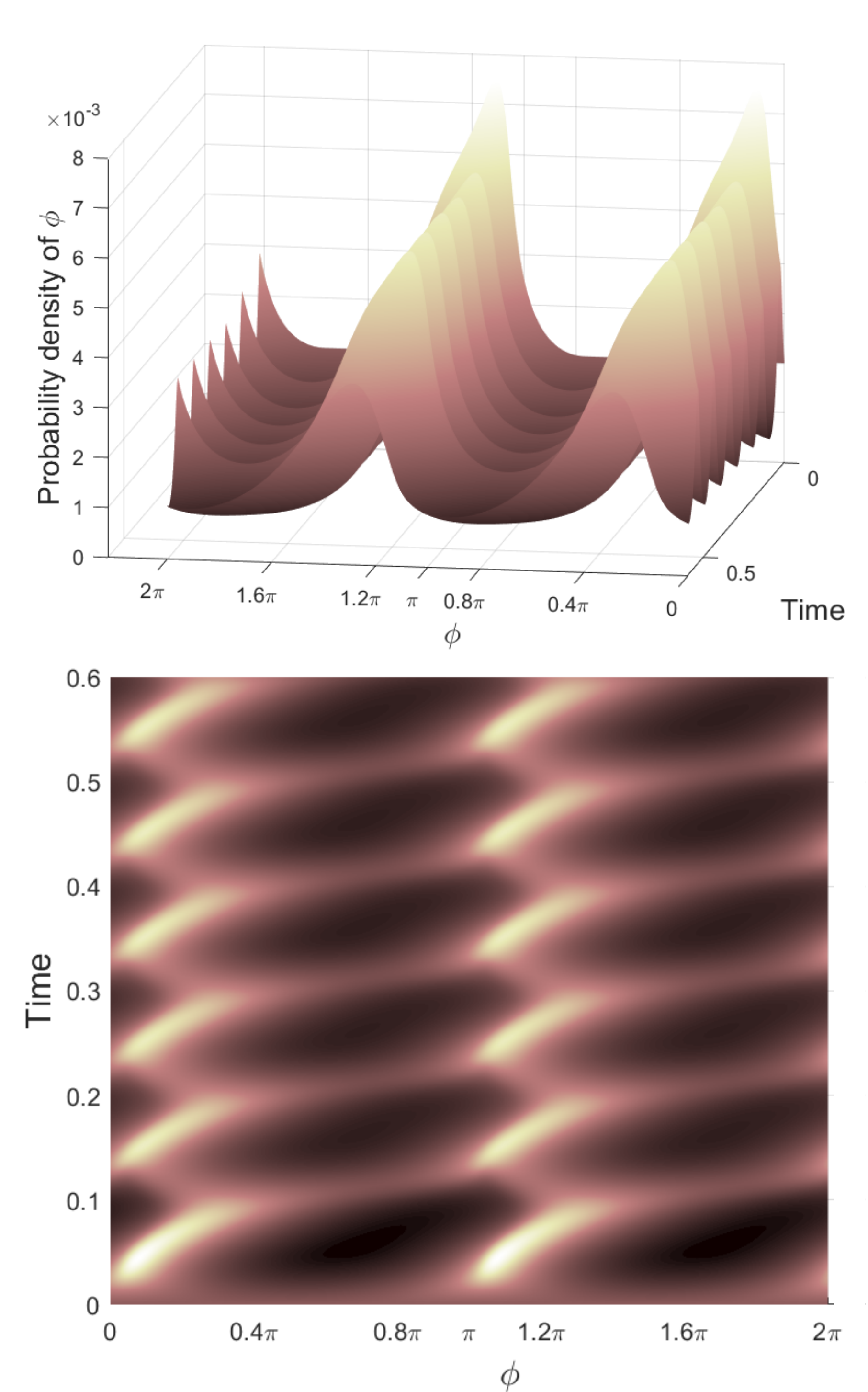}
\par\end{centering}
\caption{Probability density function for $\phi$ against time (side and top
views) brought about by competition between a time-dependent measurement
strength $\alpha_{x}=2+\sin20\pi t$ that draws the system towards
eigenstates of $\sigma_{x}$ at $\phi=0$ and $\pi$, and Rabi oscillations
characterised by $\epsilon=10$ that favour a positive drift for $\phi$.
\label{fig:pdfs}}
\end{figure}

We solve the Fokker-Planck equation for $\epsilon=10$ and a range
of values of $\alpha_{x}$ (with $\alpha_{y}=0$) to obtain stationary
pdfs $p_{{\rm st}}^{\alpha_{x}}(\phi)$. We then introduce a time-dependent
measurement strength $\alpha_{x}=2+\sin20\pi t$ to obtain a time-dependent
pdf $p(\phi,t)$ that settles into a periodic stationary state, as
shown in Figure \ref{fig:pdfs}. The principal feature to notice is
that the system is periodically attracted, statistically speaking,
towards the eigenstates of the $\sigma_{x}$ observable at $\phi=0$
and $\pi$, though displaced to higher values by the Rabi rotation.

\begin{figure}
\begin{centering}
\includegraphics[width=1\columnwidth]{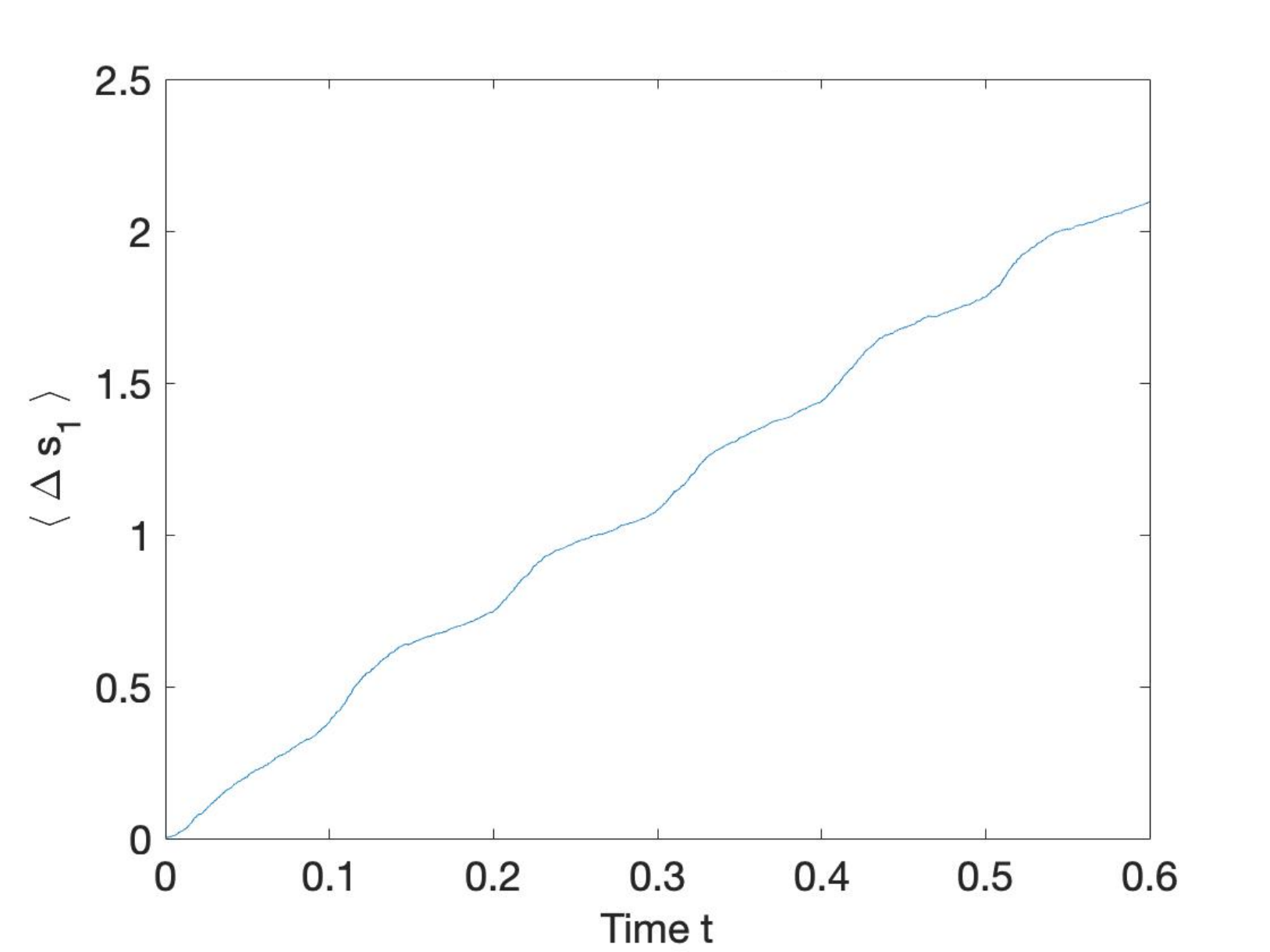}
\par\end{centering}
\caption{The average of $\Delta s_{1}$, the nonadiabatic component of stochastic
entropy production, evolving with time, for $\alpha_{x}=2+\sin20\pi t$,
$\alpha_{y}=0$ and $\epsilon=10$, adopting a periodic stationary
state to accompany the Zeno slowing down of the Rabi oscillations
and reflecting the time-dependence of the measurement strength $\alpha_{x}$.
\label{fig:mean-stochastic-entropy-1}}
\end{figure}

We have calculated the average of $\Delta s_{1}$ as a function of
time for $\epsilon=10$ and $\alpha_{x}=2+\sin20\pi t$ using methods
described in \citep{SpinneyFord12b} and the results are given in
Figure \ref{fig:mean-stochastic-entropy-1}. Since the system is prevented
from reaching a stationary state through the time-dependence of the
measurement strength, the mean rate of change of this component of
stochastic entropy production never falls to zero, but instead continues
to evolve periodically.

\begin{figure}
\begin{centering}
\includegraphics[width=1\columnwidth]{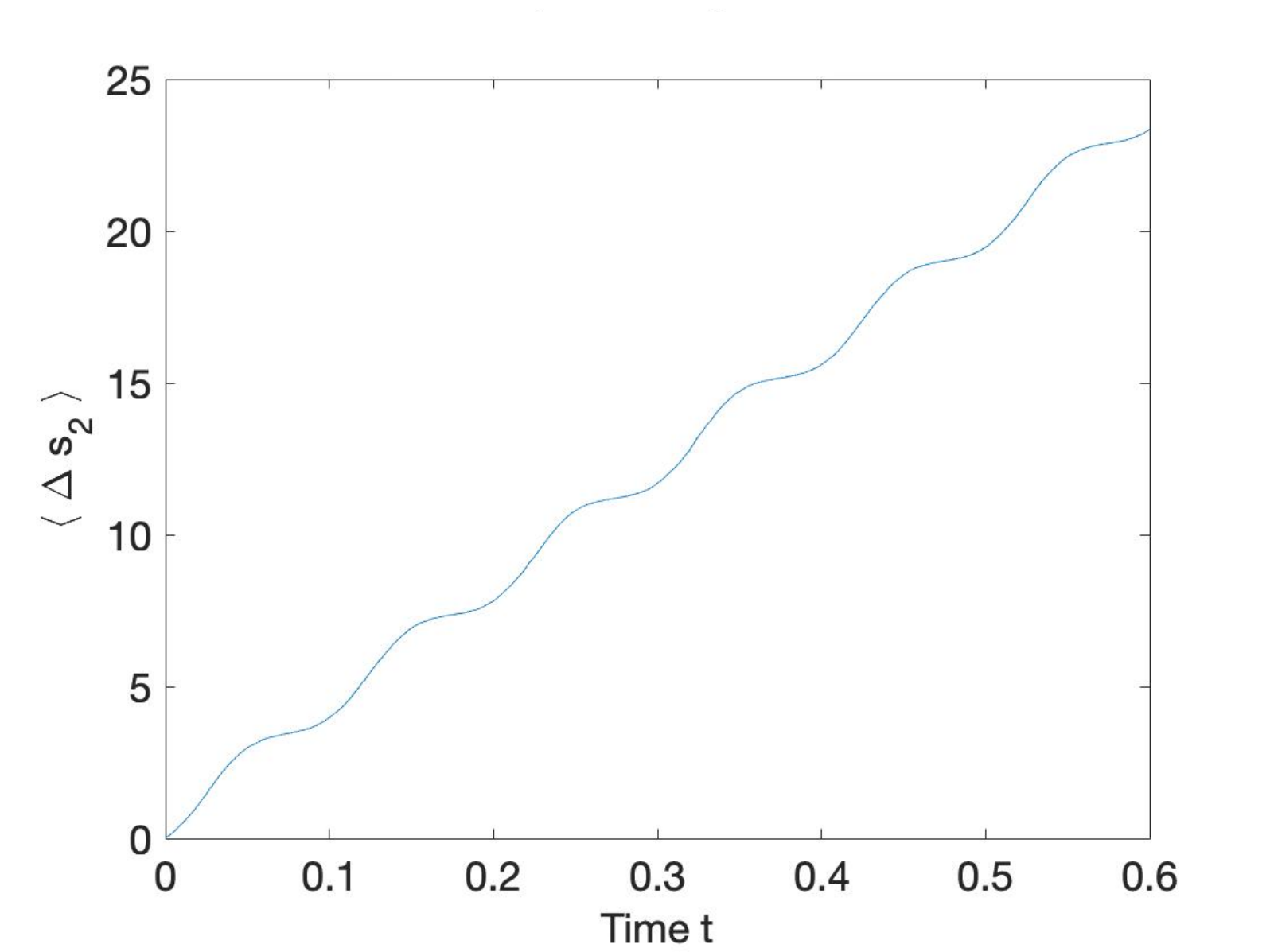}
\par\end{centering}
\caption{Average of $\Delta s_{2}$, the adiabatic component of stochastic
entropy production, against time, for $\alpha_{x}=2+\sin20\pi t$,
$\alpha_{y}=0$ and $\epsilon=10$. \label{fig:mean-stochastic-entropy-2}}
\end{figure}

The average of the second component of stochastic entropy production
evolves according to \citep{SpinneyFord12a}
\begin{equation}
\frac{d\langle\Delta s_{2}\rangle}{dt}=\int\frac{p}{D}\left(\frac{J_{{\rm st}}^{{\rm irr}}(\phi^{{\rm T}})}{p_{{\rm st}}^{\alpha_{x}}(\phi^{{\rm T}})}\right)^{2}d\phi,\label{eq: 15}
\end{equation}
where $\phi^{{\rm T}}$ is the transform of $\phi$ under time reversal:
since $\phi$ is odd, $\phi^{{\rm T}}=-\phi$. $\Delta s_{2}$ is
a contribution to stochastic entropy production arising from the breakage
of detailed balance, which permits the emergence of a non-zero irreversible
probability current in a stationary state, given by
\begin{equation}
J_{{\rm st}}^{{\rm irr}}(\phi)=A^{{\rm irr}}p_{{\rm st}}^{\alpha_{x}}(\phi)-\frac{\partial}{\partial\phi}D(\phi,\alpha_{x})p_{{\rm st}}^{\alpha_{x}}(\phi),\label{eq: 16}
\end{equation}
where the diffusion coefficient is specified by the current value
of $\alpha_{x}$. Esposito and Van den Broeck referred to $\Delta s_{2}$
as the adiabatic entropy production \citep{adiabaticnonadiabatic0}
and Spinney and Ford, who included a consideration of dynamical variables
that are odd as well as even under time reversal symmetry, denoted
it the generalised housekeeping entropy production \citep{SpinneyFord12a}.
Like the nonadiabatic entropy production, its mean rate of change
is never negative. The evolution of $\langle\Delta s_{2}\rangle$
for $\epsilon=10$ and $\alpha_{x}=2+\sin20\pi t$ is illustrated
in Figure \ref{fig:mean-stochastic-entropy-2}.

\begin{figure}
\begin{centering}
\includegraphics[width=1\columnwidth]{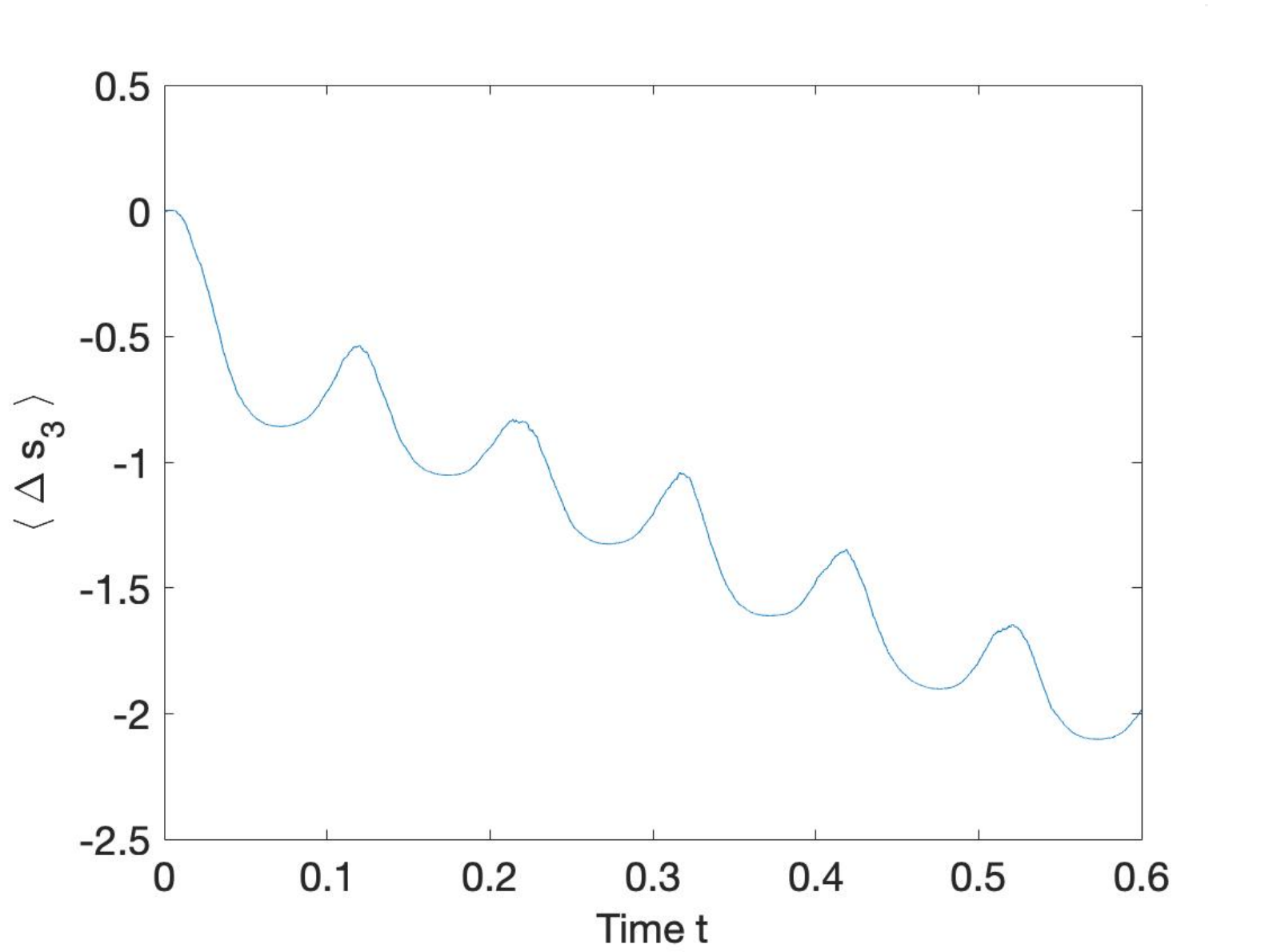}
\par\end{centering}
\caption{Average of $\Delta s_{3}$, the transient housekeeping stochastic
entropy production, against time, for $\alpha_{x}=2+\sin20\pi t$,
$\alpha_{y}=0$ and $\epsilon=10$. \label{fig:mean-stochastic-entropy-3}}
\end{figure}

The average rate of change of the third contribution to the stochastic
entropy production may be written
\begin{equation}
\frac{d\langle\Delta s_{3}\rangle}{dt}=-\int\frac{\partial p}{\partial t}\ln\frac{p_{{\rm st}}^{\alpha_{x}}(\phi)}{p_{{\rm st}}^{\alpha_{x}}(\phi^{{\rm T}})}d\phi.\label{eq: 17}
\end{equation}
$\Delta s_{3}$ is a contribution associated with relaxation towards
a stationary state and in this respect is similar to $\Delta s_{1}$.
It explicitly vanishes when there are no odd variables in the dynamics,
but here it does not vanish. It was designated the transient housekeeping
entropy production by Spinney and Ford \citep{SpinneyFord12a}. The
evolution of $\langle\Delta s_{3}\rangle$ for $\epsilon=10$ and
$\alpha_{x}=2+\sin20\pi t$ is illustrated in Figure \ref{fig:mean-stochastic-entropy-3}.
Notice that a negative mean rate of production is permitted, in contrast
to the other two contributions. The mean rate of change of $\Delta s_{{\rm tot}}$
is, of course, positive for all times, in accordance with the second
law \citep{seifert2008stochastic}.

If we were to re-introduce the measurement of $\sigma_{y}$, and hence
create conditions for an anti-Zeno effect in the dynamics, the stochastic
entropy production would similarly divide into three components and
quantify the relative contributions of different sources of irreversibility.

\section{Conclusions\label{sec:Conclusions}}

Employing the framework of quantum state diffusion as a model of the
evolution of an open quantum system allows us to investigate the effect
of measurement on the intrinsic dynamics of a quantum system. We have
previously investigated a Zeno effect in a multi-level bosonic system:
a slowing down of Rabi oscillations, on average, when measurements
are performed to determine the level currently occupied \citep{Walls22}.
Here we extend that study to demonstrate that simultaneous measurement
of a second, non-commuting observable can produce a counter-intuitive
anti-Zeno effect, specifically that the slowed down evolution can
be speeded up when measurement of the second observable is introduced.

The principal aim of the study, however, has been to compute the stochastic
entropy production associated with the evolution of a two-level system
when a Zeno effect is operating as a result of the continuous measurement
of one observable. Our investigation of the division of the stochastic
entropy production into its three components is motivated by a wish
to demonstrate that the ideas underpinning stochastic thermodynamics
can apply with equal validity to classical and quantum mechanics.
Stochastic entropy production measures the irreversibility of the
evolution of a system when subjected to unpredictable external disturbance.
This is the extent to which two sequences of events, one the reverse
of the other, occur with different probabilities in such circumstances.
We take the trajectory followed by the reduced density matrix of an
open quantum system, when it is subjected to continuous measurement,
to be analogous to the Brownian path of a classical particle under
the influence of an unpredictable environment. Irreversibility occurs
in both situations and can be quantified.

The division of stochastic entropy production into components demonstrates
how irreversibility can be associated with the relaxation of a system
towards stationarity (components $\Delta s_{1}$ and $\Delta s_{3}$)
and with the breakage of detailed balance and the consequent existence
of a nonequilibrium stationary state (component $\Delta s_{2}$).
These three contributions emerge in the two-level quantum system when
we make the strength of measurement a periodic function of time, such
that the reduced density matrix and the mean stochastic entropy production
also evolve periodically. We conclude that stochastic entropy production
associated with nonequilibrium behaviour, reflecting a continuing
loss of information concerning the configuration of the world, can
be demonstrated in open quantum systems as well as in classical situations.
\begin{acknowledgments}
We thank Haocheng Qian for assistance in the early stages of this
work.
\end{acknowledgments}

\bibliography{ref}

\end{document}